\documentclass[aps,prl,reprint,groupedaddress]{revtex4-2}

\usepackage{upgreek}
\usepackage{color}
\usepackage{amsmath}
\usepackage{tipa}
\usepackage{bbm}
\usepackage{txfonts}
\usepackage{graphicx}   
\usepackage{dcolumn}    
\usepackage{bm}         
\usepackage{amssymb}    
\usepackage{latexsym}   
\usepackage{braket}
\usepackage{hyperref}
\usepackage{multirow}

\hypersetup{hypertex=true,colorlinks=true,linkcolor=blue,anchorcolor=blue,citecolor=blue}
\begin{document}

\title{de-Broglie Wavelength Enhanced Weak Equivalence Principle Test for Atoms in Different Hyperfine States}
\author{Yao-Yao Xu}
\author{Xiao-Bing Deng}
\author{Xiao-Chun Duan}
\email[]{duanxiaochun2011@hust.edu.cn}
\author{Lu-Shuai Cao}
\author{Min-Kang Zhou}
\author{Cheng-Gang Shao}
\author{Zhong-Kun Hu}
\email[]{zkhu@hust.edu.cn}
\affiliation{MOE Key Laboratory of Fundamental Quantities Measurement, Hubei Key Laboratory of Gravitation and Quantum Physics, School of Physics, Huazhong University of Science and Technology, Wuhan 430074, China}

\date{\today}

\begin{abstract}
We report a hyperfine-states related weak equivalence principle (WEP) test which searches for possible WEP violation signal in single atom interferometer. 
With the ground hyperfine states $\left|F=1\right\rangle$ and $\left|F=2\right\rangle$ of $^{87}$Rb atoms simultaneously scanned over different paths in a Raman Mach-Zehnder interferometer (MZI), the difference of the free fall accelerations for the atom in the two hyperfine states is encoded into the phase shift of the MZI, contributing a WEP test signal. 
The test signal can be extracted out by reversing the direction of the effective wave vector of the Raman laser to suppress direction-dependent disturbances.
More importantly, de-Broglie wavelength of cold atoms can be utilized to enhance the test signal in our scheme, which helps to improve the upper bound of the WEP test for atoms in different hyperfine states to $2.9\times10^{-11}$, about one order of magnitude lower than the previous record.
\end{abstract}

\maketitle

The weak equivalence principle (WEP) specifies the equivalence of inertia and weight for test masses. It is one of the fundamental assumptions for Einstein’s general relativity \cite{Weinberg1972-WEIGAC,Charles1973Gravitation}, and has profound implications for understanding gravitational interaction. Despite the great success of Einstein’s theory, this assumption still needs experimental verification, not to mention that many candidate unification models require the violation of WEP \cite{Damour_2012}. Traditional WEP tests explore macroscopic objects of different compositions \cite{PhysRevD.50.3614,PhysRevLett.59.609,PhysRevLett.62.1941,PhysRevLett.69.1722,Dickey482,PhysRevLett.93.261101,PhysRevLett.100.041101,PhysRevLett.119.231101}, and have achieved a level of $10^{-13}$ to $10^{-14}$ \cite{PhysRevLett.119.231101}. There are also tests searching for possible state-related violation of WEP using macroscopic objects, such as polarized bodies \cite{Hou2000Rotatable}, rotating gyroscopes \cite{PhysRevLett.63.2701,PhysRevLett.64.825,PhysRevLett.64.2115,PhysRevD.66.022002,PhysRevD.65.042005,PhysRevLett.107.051103} and chiralities \cite{PhysRevLett.121.261101}.

\begin{table}[htbp]\centering
\caption{WEP tests using atoms.}\label{table1}
\footnotesize
\begin{tabular}{ccccc}

\hline
\hline
test mass A&test mass B&$\eta$&  &Ref\\
\hline
Cs&corner cube&7(7)$\times$10$^{-9}$& &\cite{Peters1999measurement}\\
Rb&corner cube&$4.3(6.4)\times 10^{-9}$& &\cite{Merlet_2010}\\
Sr&corner cube&$1.6(1.4)\times 10^{-7}$& &\cite{PhysRevLett.106.038501}\\

$^{87}$Rb&$^{39}$K&$0.3(5.4)\times 10^{-7}$& &\cite{PhysRevLett.112.203002}\\
$^{87}$Rb&$^{39}$K&$0.9(3.0)\times 10^{-4}$& &\cite{Barrett_2016}\\
$^{87}$Rb&$^{39}$K&$-1.9(3.2)\times 10^{-7}$& &\cite{Albers2020Quantum}\\

$^{85}$Rb&$^{87}$Rb&$1.2(1.7)\times 10^{-7}$& &\cite{PhysRevLett.93.240404}\\
$^{87}$Rb&$^{85}$Rb&$1.2(3.2)\times 10^{-7}$& &\cite{PhysRevA.88.043615}\\
$^{88}$Sr&$^{87}$Sr&$0.2(1.6)\times 10^{-7}$& &\cite{PhysRevLett.113.023005}\\
$^{85}$Rb&$^{87}$Rb&$2.8(3.0)\times 10^{-8}$& &\cite{PhysRevLett.115.013004}\\
$^{87}$Rb&$^{85}$Rb&$1.6(3.8)\times 10^{-12}$& &\cite{PhysRevLett.125.191101}\\
$^{87}$Rb&$^{85}$Rb&$-0.8(1.4)\times 10^{-10}$& &\cite{PhysRevA.104.022822}\\

$^{85}$Rb,$F$=3&$^{85}$Rb,$F$=2&$0.4(1.2)\times 10^{-7}$& &\cite{PhysRevLett.93.240404}\\
$^{87}$Rb,$m_F=+1$&$^{87}$Rb,$m_F=-1$&$0.2(1.2)\times 10^{-7}$& &\cite{PhysRevLett.117.023001}\\
$^{87}$Rb,$F$=1&$^{87}$Rb,$F$=2&$1.0(1.4)\times 10^{-9}$& &\cite{Rosi_2017}\\
$^{87}$Rb,$F$=1&$^{87}$Rb,$F$=1$\oplus$2&$3.3(2.9)\times 10^{-9}$& &\cite{Rosi_2017}\\
$^{87}$Rb,$F$=1&$^{87}$Rb,$F$=2&$0.9(2.7)\times 10^{-10}$& &\cite{KeZhang43701}\\
$^{87}$Rb,$F$=1&$^{87}$Rb,$F$=2&$0.9(2.9)\times 10^{-11}$& &this work\\

\hline
\hline
\end{tabular}\\

\end{table}
\normalsize

Along with the development of matter-wave interferometry, WEP tests have been extended to microscopic domain, as listed in Table \ref{table1}. In analogy with macroscopic tests using objects of different compositions, WEP tests have been performed by comparing the free fall accelerations between macroscopic objects and microscopic particles \cite{PhysRevLett.34.1472,Peters1999measurement,Merlet_2010,PhysRevLett.106.038501} or between different microscopic particles \cite{Albers2020Quantum,PhysRevLett.93.240404,PhysRevA.88.043615,PhysRevLett.112.203002,PhysRevLett.115.013004,PhysRevLett.113.023005,Kuhn_2014,Barrett_2016,PhysRevLett.117.023001,Rosi_2017,PhysRevLett.125.191101,KeZhang43701,PhysRevA.104.022822}.
A level of $10^{-12}$ has been achieved for this kind of tests \cite{PhysRevLett.125.191101}. There are also special WEP tests that can be uniquely carried out in the microscopic domain, for example, atoms in different hyperfine states (HSs) \cite{Rosi_2017,KeZhang43701,PhysRevA.104.022822}, bosonic particles versus fermionic particles \cite{PhysRevLett.113.023005}, and atoms in different spin orientations \cite{PhysRevLett.117.023001}. Test levels of this kind range from $10^{-4}$ to $10^{-10}$. 

\begin{figure}
\includegraphics[width=8 cm]{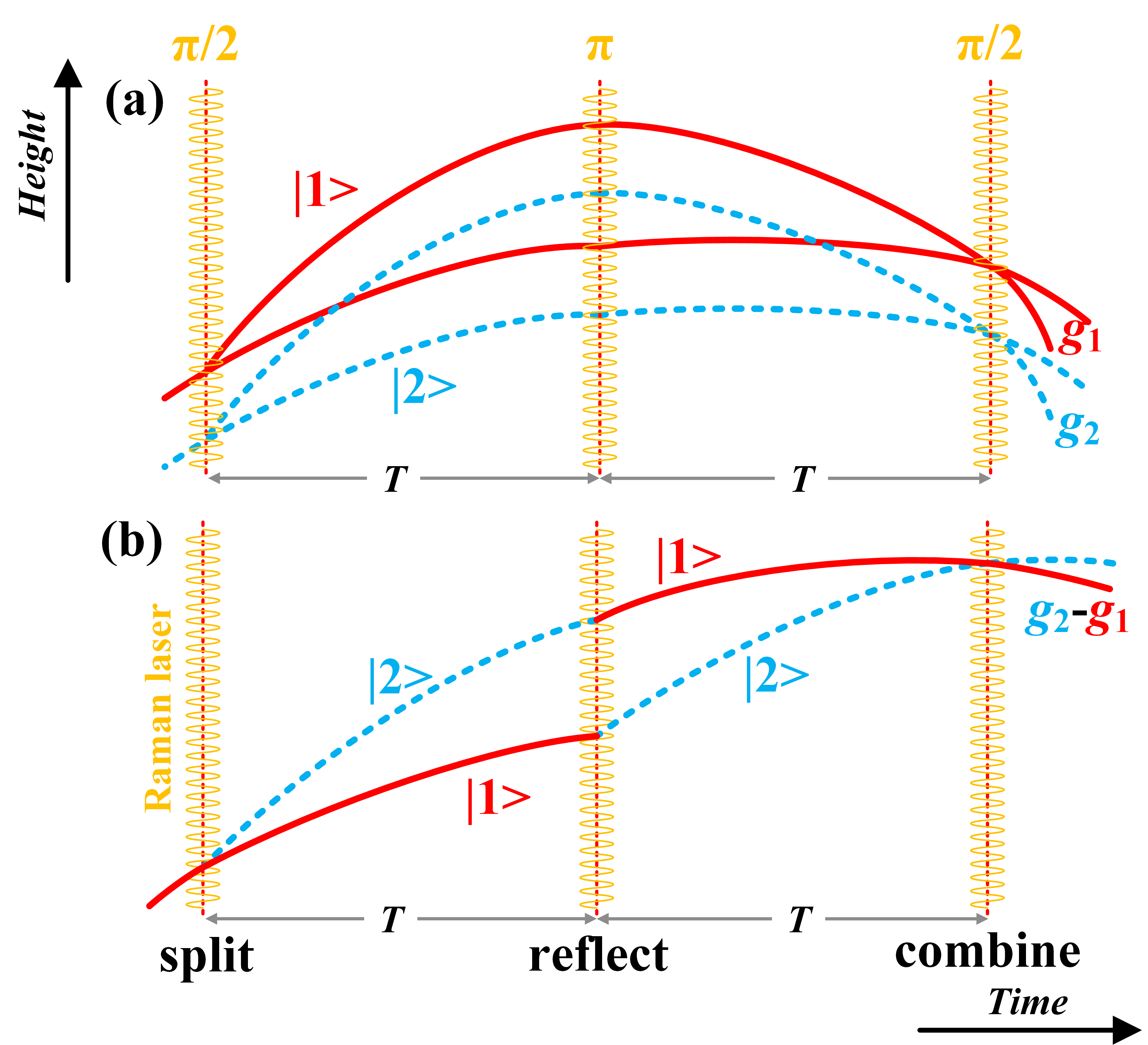}
\caption{
Typical MZIs where the wave packet is split, reflected and combined sequentially by Raman laser pulses. The top figure shows the usual WEP tests scheme where the HS of the atom is invariant during interference and the free fall accelerations are measured separately by two interferometers. And the bottom figure shows our WEP test scheme where the interferometer scans the two HSs simultaneously in one interferometer.
$g_i$ ($i$=1,2) stands for the free fall accelerations in the corresponding HS.
}
\label{typical}
\end{figure}

WEP tests using atoms, however, usually follow the classical strategy of gravity measurements, namely measuring the free fall accelerations of concerned test masses separately and then comparing them to give the test signal, as shown in Fig. \ref{typical}(a). 
Moreover, usual WEP tests with atom interferometers (AIs) have not made full use of the short de-Broglie wavelength of cold atoms, which is a key advantage of AIs over optical interferometers \cite{Baudon_1999}.
Recently, there has arisen interest in exploring superposition states of atoms \cite{Rosi_2017,PhysRevResearch.2.043240,PhysRevX.10.021014}. For instance, Giulini has discussed possible WEP violation in single AI \cite{GiuliniEquivalence}. Inspired by these works, we perform a WEP test allowing direct access to the free fall acceleration difference of atoms in different HSs in single AI. 
The WEP test adapts a Raman Mach-Zehnder interferometer (MZI) with atomic wave packet evolving in different HSs for different interferometer branches, namely simultaneously scanning the two HSs under test for single interferometer, as shown in Fig. \ref{typical}(b).
The free fall acceleration difference for atoms in the two HSs is encoded to the phase difference of the two paths, contributing a WEP test signal. 
This WEP test signal can be extracted out by reversing the direction of the effective wave vector of Raman laser to suppress direction-dependent disturbances. More importantly, as we will demonstrate, the test signal here is dependent on the de-Broglie wavelength of the cold atoms, which manifests as an significant improvement to usual WEP tests with atoms. Benefiting from the enhancement of the test signal by de-Broglie wavelength, a precision of $2.9\times 10^{-11}$ is achieved for our WEP test with respect to atoms in different HSs, almost one order of magnitude improvement over the previous tests.

In our WEP test experiment, we perform the AI measurements with $^{87}\rm{Rb}$ atoms in two magnetic-insensitive sublevels $\left|1\right\rangle =\left|F=1,m_F=0\right\rangle$ and $\left|2\right\rangle=\left|F=2,m_F=0\right\rangle$.
The measurement process is as follows: initially the atoms are prepared in $\left|1\right\rangle$, and a $\pi/2-\pi-\pi/2$ Raman-pulse train sends the atom wavepacket to two interference paths, which merges at the end of the interference.
The Raman pulses not only manipulate the paths, but also flips the HSs.
In order to describe the two HSs of $^{87}$Rb atoms simultaneously involved during the interference in single AI,
quantum formulation of the inertial mass $\widehat{M}_i$ and the gravitational mass $\widehat{M}_g$ are introduced as \cite{Zych2018Quantumformulation,Zych2011interferometric,Pikovski2015decoherence,ZychSystem,Rosi_2017}
\begin{equation}\label{intmass}
\widehat M_{\alpha}=m_{\rm{\alpha}}\hat I+\frac{\widehat H_{\alpha}}{c^2}, 
\end{equation}
where the subscript $\rm{\alpha}=\rm{i,g}$ refers to inertia and weight, respectively, and $\widehat H_{\alpha}$ is the Hamiltonian describing the internal interaction of the atom.
In Eq. (\ref{intmass}), $\widehat H_{\alpha}/{c^2}$ stands for the contribution of the internal energy to mass. The quantum formulation of the WEP is then $\widehat{M}_{\rm g} \widehat{M}_{\rm i}^{-1}=\hat{I}$. Only the diagonal elements of $\widehat{M}_{\rm g} \widehat{M}_{\rm i}^{-1}$ are concerned here.
The center of mass acceleration for atoms in each HS can be expressed as (see the supplementary material)
\begin{equation}\label{atomaccel}
a_j=\left\langle j\left|\widehat{M}_{\rm g} \widehat{M}_{\rm i}^{-1}\right|j\right\rangle g=\frac{m_{\rm g}}{m_{\rm i}}r_j g,
\end{equation}
where the subscript $j=1,2$ specifies the corresponding HS, and $g$ stands for the strength of the local gravitational field.
$r_j$ is the parameter which quantifies the difference of the contribution of internal energy to inertia and weight. $r_j$ can also account for other violation mechanisms which cause different free fall accelerations for atoms in different HSs \cite{PhysRevLett.67.1735}. Further calculation yields the E$\rm \ddot o$tv$\rm \ddot o$s parameter for our WEP test to be $\eta =r_2-r_1$.
In addition to an approximation of only keeping the lowest order in $1/c^2$ during the calculation of $\widehat{M}_{\rm g} \widehat{M}_{\rm i}^{-1}$ as in Ref \cite{Rosi_2017}, here the off-diagonal elements of $\widehat{M}_{\rm g} \widehat{M}_{\rm i}^{-1}$ are further assumed to be zero.

After the introduction of hyperfine-state-dependent free-fall acceleration, the two branches of the interferometer don't close any more, resulting in a separation $\Delta z=(r_1-r_2)(m_{\rm g}/m_{\rm i})gT^2$ at the exit of the interferometer \cite{GiuliniEquivalence}. In this situation, the total phase shift, including the phase shift due to free evolution between the Raman pulses, the phase shift due to the interaction with the Raman pulses, and the phase shift due to the separation, is calculated to be
\begin{equation}\label{viophase}
\Delta \phi =-(r_2-r_1)(\vec k_m^{\rm eff}\cdot \hat g)\left(\frac{m_{\rm g}}{m_{\rm i}}\right)gT^2+r_1(\vec k_L^{\rm eff}\cdot \hat g)\left(\frac{m_{\rm g}}{m_{\rm i}}\right)gT^2,
\end{equation}
where $\hat g$ stands for the direction of local gravitational acceleration. $\vec k_L^{\rm eff}=\vec k_1-\vec k_2$ is the effective wave vector of Raman laser. $\vec k_m^{\rm eff}=m_{\rm i}\vec V_{\pi}/\hbar$ stands for the wave number of matter wave,
and $\vec V_{\pi}$ is the velocity of the atom in the state $\left|1\right\rangle$ at the reflecting $\pi$ pulse.
It is clearly shown in Eq. (\ref{viophase}) that the interferometer includes a phase shift sensitive to the WEP test signal $r_2-r_1$, indicating that a WEP test signal indeed exists in single AI with two HSs simultaneously scanned.
More importantly, the test signal is proportional to the wave number of de-Broglie wave.
Given that the de-Broglie wavelength can be much shorter than light wavelength,
this AI test scheme is expected to improve the precision of the WEP test.
During the calculation of $\Delta \phi$, both the free evolution phase shift and the separation phase shift are related to the de-Broglie wavelength, contributing to the WEP test signal. 
For comparison, in a Raman MZI with a linear gravity gradient present, the free evolution phase shift and the separation phase shift nearly cancel, although the two phase shifts are both related to the de-Broglie wavelength individually.
Therefore, the incorporation of WEP test in a Raman MZI with two HSs involved leads to the de-Broglie wavelength related phase shift.

According to Eq. (\ref{viophase}), the WEP test signal emerges in single interferometer. However, the second term, induced by normal gravity $\left({m_{\rm g}}/{m_{\rm i}}\right)g$, is still the major one among the total phase shift. A common-mode measurement by reversing $\hat k_L^{\rm eff}$ (the direction of $\vec k_L^{\rm eff}$) can be applied to suppress the phase shift due to normal gravity and extract out the WEP test signal from the total phase shift.
Denoting $k_L^{+}(k_L^-)$ as the magnitude of Raman laser wave vector at $\pi$ pulse for $\hat k_L^{\rm eff}$ along (against) the direction of local gravity, the common-mode measurement 
is to add the phase shifts of the interferometers with opposite directions of $\vec k_L^{\rm eff}$.
The common-mode measurement result $\Delta \Phi_{\rm c}$ is expressed as
\begin{equation}\label{commonphase}
\Delta \Phi_{\rm c}\! \equiv \frac{\Delta \phi^+\!+\Delta \phi^-}{2}\! =-(r_2-r_1)(\vec k_m^{\rm eff}\cdot \hat g)\left(\frac{m_{\rm g}}{m_{\rm i}}\right)gT^2+r_1\frac{k_L^{+}\! -k_L^{-}}{2}\left(\frac{m_{\rm g}}{m_{\rm i}}\right)gT^2,
\end{equation}
where $\Delta \phi^{\pm}$ is the phase shift of the interferometer with corresponding $\hat k_L^{\rm eff}$. 
Given that $k_L^{+}\approx k_L^{-}$, the second term in Eq. (\ref{commonphase}) vanishes, and the WEP test signal can be extracted out.
The sensitive axis of the atom gravimeter is determined by $\hat k_L^{\rm eff}$, and thus phase shifts due to nomal gravity cancel in the common-mode measurement with $\hat k_L^{\rm eff}$ reversed.
However, the measurand of our WEP test is the difference of the free fall accelerations rather than the acceleration itself, and the WEP test signal survives in the common-mode measurement.
Compared with usual WEP tests shown in Fig. \ref{typical}(a), our proposed test scheme shown in Fig. \ref{typical}(b) gives the test signal directly in single Raman MZI. 
Moreover, the WEP test signal is dependent on the de-Broglie wavelength, which is expected to greatly improve the test precision.
And this WEP test signal can be extracted out by the common-mode measurement.

We perform the novel WEP test on the atom gravimeter as described in Ref. \cite{Xu2018onsite,Cui2018time}. 
In our experiment, about $10^9$ atoms are cooled and trapped in a magneto-optical trap with a loading time of 300 ms, and launched upwards subsequently with an aimed height of 0.6 m. 
The temperature of the cloud after launch is about 4 $\mu$K.
During the flight, the atoms are firstly state prepared with a Raman $\pi$ pulse and then subject to the $\pi/2-\pi-\pi/2$ Raman-pulse sequence in a magnetically shielded interferometry tube. A coil wrapped around the tube with an injection current of 15 mA supplies a bias magnetic field of about 180 mG. After the interference, the atoms are detected with a normalization detection. It takes one second to complete the whole process.
In comparison to usual gravity measurements by Raman MZIs, there are two differences for the proposed WEP test experiment. 
Firstly, for the usual gravity measurements by MZI, atoms interact with the interfering Raman $\pi$ pulse near the apex during the flight \cite{PhysRevLett.117.023001}.
Disturbances from magnetic field inhomogeneity, light shift, etc. can be suppressed with this configuration. However, the corresponding velocity of the atomic cloud at $\pi$ pulse approaches zero in that situation, resulting in a nearly zero value for $k_m^{\rm eff}$. Thus for the proposed WEP test here, the interfering Raman $\pi$ pulse is applied when the atom cloud is far away from the apex, obtaining a large value of $V_{\pi}$.
A typical value of about 1 m/s for $V_{\pi}$ is chosen here, which allows an interrogation time of 100 ms for the 0.6 m height fountain in our instrument. 
This $V_{\pi}$ corresponds to a value of $1.4\times 10^9$ $\rm m^{-1}$ for $k_m^{\rm eff}$, larger than typical wave number of Raman laser by about eighty times. 
Secondly, for usual gravity measurements, the differential-mode measurement by reversing $\hat k_L^{\rm eff}$ is explored to extract the gravity induced phase shift, which largely suppresses $\vec k_L^{\rm eff}$ independent disturbances, such as the second-order Zeeman shift and ac-Stark shift. 
However, for the proposed WEP test, normal gravity becomes a disturbance, since the measurand here is the difference of the free fall accelerations rather than the gravitational acceleration itself. 
The common-mode measurement helps to suppress the phase shift due to normal gravity and thus extract out the test signal.

\begin{figure}
\includegraphics[width=8 cm]{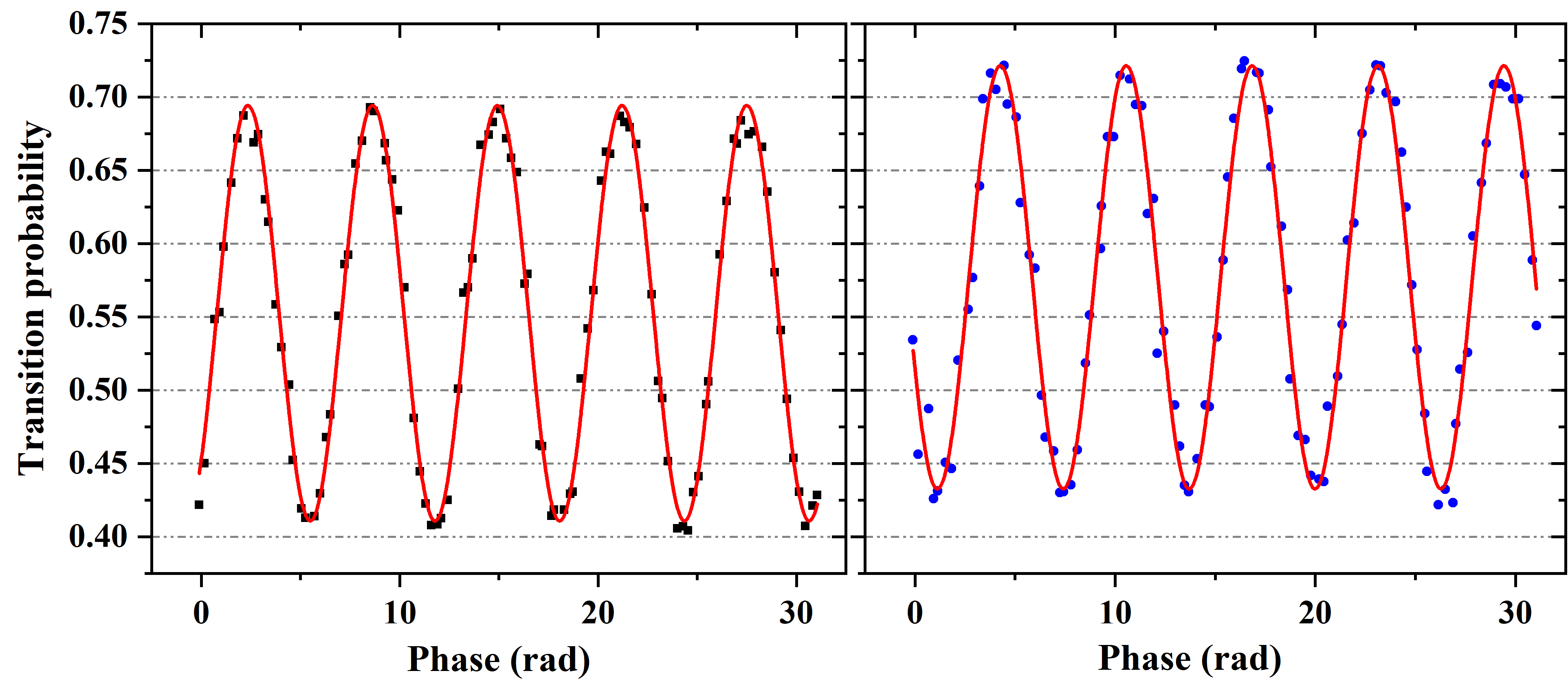}
\caption{Typical interference fringes for different directions of $\vec k_L^{\rm eff}$. The direction of $\vec k_L^{\rm eff}$ is reversed every one hundred shots, and the corresponding fringes are displayed as black squares and blue dots, respectively. The solid red lines indicate least-square fits to sinusoidal functions.
The fringes for different directions of $\vec k_L^{\rm eff}$ shift a bit mainly due to imperfection normalization detection in the experiment.}
\label{typicalfringes}
\end{figure}
\begin{table}[!hbp]
\tabcolsep 3mm \caption{\label{errortable}Main systematic errors affecting the common-mode measurement for the WEP test.}
\begin{center}
\setlength{\tabcolsep}{5mm}{
\begin{tabular}{ccc}
\hline
\hline
Systematic effect&Bias/mrad & Un./mrad \\
\hline
Magnetic field gradient&192 &3   \\
Single-photon light shift &4   &2   \\
Two-photon light shift &-0.8 &0.3 \\
Finite light speed &-11.2   &0.1   \\
Frequency chirp &-5.6   &0.1   \\
Gravity gradient &0.3 &0.1 \\
\hline
Total&178.7   &3.6   \\
\hline
$\Delta \Phi_{\rm c}$&180 &2 \\
\hline
Result&1.3   &4.1    \\
\hline
\hline
\end{tabular}}
\end{center}
\end{table}

We obtain one fringe for every twenty shots by scanning the chirp rate of the Raman laser' effective frequency step by step, with the direction of $\vec k_L^{\rm eff}$ reversed for every five fringes.
Typical fringes are shown in Fig. \ref{typicalfringes}.
One-hour data acquisition gives a result of 180(2) mrad for the common-mode result $\Delta \Phi _{\rm c}$, where the quoted uncertainty comes from the statistics error. 

The evaluation of the main systematic errors is summarized in Table \ref{errortable}. The disturbances related only to the external freedoms of atoms, such as Coriolis effect and normal gravity, are suppressed by the common-mode measurement. 
$k_{L}^{+}$ and $k_{L}^{-}$ are different at the $\pi$ pulse due to non-zero value of $V_{\pi}$.
The difference $\left|k_L^+-k_L^-\right|/k_L^+$ is about $7\times10^{-9}$, giving a suppression ratio at level of $10^{-9}$ for normal gravity in the common-mode measurement.
The residual contribution of normal gravity is -5.6(1) mrad. For the disturbances related to the wavefront of the Raman laser, for example, the wavefront abbreviations \cite{Louchet_Chauvet_2011}, the corresponding error is suppressed by $\left| {{k_1} - {k_2}} \right|/k^{\rm eff}_L\sim 10^{-5}$ as different beams travel through additional optical elements for different directions of $\vec k_L^{\rm eff}$. Here, $k_{1,2}$ denotes the corresponding wave number of the two beams making up the Raman laser. 
Therefore the influence of wavefront abbreviations can be safely neglected.
For the light shifts induced by the Raman laser, single-photon light shift (SPLS) and two-photon light shift (TPLS) are considered. 
SPLS induced by the Raman laser is measured by the microwave Rabi spectroscopy \cite{PhysRevA.93.053615}. The intensity ratio of the two beams comprising of the Raman laser is varied to induce a modulation of the light shift, and the corresponding variation of the interferometer phase shift is measured. The slope of the phase shift varying with the light shift is obtained from this modulation experiment, which is -0.207(6) mrad/Hz. Correspondingly, the determined contribution due to light shift is 4(2) mrad.
TPLS is mainly induced by the off-resonant counterpropagating Raman beams here, and the shift is 
$\delta \omega_{\rm TPLS}=-\Omega _{{\rm{eff}}}^2/8(\vec k_L^{\rm eff}\cdot \vec V_{\pi /2}) - \Omega _{{\rm{eff}}}^2/8(\vec k_L^{\rm eff}\cdot \vec V_{\pi /2}+2{\omega _r})$ \cite{PhysRevA.78.043615}.
Here $\Omega _{{\rm{eff}}}=2\pi \times 10.0(4)$ kHz is the effective two-photon Rabi frequency, ${\omega _r} = \hbar {(k_L^{{\rm{eff}}})}^2/2m$ is the recoil frequency, and $V_{\pi /2}$=0.085(1) m/s is the velocity of the atom cloud at the second interfering $\pi /2$ pulse. 
The error due to TPLS can be still partially suppressed by the common-mode measurement even with the presence of $\omega _r$. 
The resultant contribution is -0.8(3) mrad. 
The disturbances related to the internal energy shift of atoms during the free evolution of the wave packets, such as dc-Stark shift and Zeeman shift, will induce an effective differential acceleration for atoms in different HSs. For $^{87}$Rb atoms, the differential dc-Stark shift for the two HSs is quite small \cite{PhysRevA.82.022510}, and the corresponding error is safely neglected. The disturbance due to Zeeman shift can be divided into the bias current dependent and independent parts. The two parts are individually determined using different methods (see the supplemental material for detail), and the final contributions is 192(3) mrad. In addition, the influence of the finite light speed (FLS) is calculated as $-2\alpha_{\rm L}V_{\pi}T^2/c$ according to Ref. \cite{yujie2016finite}, where $\alpha_{\rm L}$ is the frequency chirp rate of the Raman laser. Taking into account the measured velocity $V_{\pi}$ of 1.064(1) m/s and the chirp rate $\alpha_{\rm L}$ of 25.10 MHz/s, the FLS induced phase shift is -11.2(1)mrad.
The gravity gradient induces a residual phase shift in the common-mode measurement, which is $k_L^{\rm eff}r_1\left(\frac{m_{\rm g}}{m_{\rm i}}\right)\gamma V_r T^3/2$ ($V_r=\hbar k_L^{{\rm{eff}}}/m$ is the recoil velocity).
Substituting with $\left(\frac{m_{\rm g}}{m_{\rm i}}\right)\gamma \thickapprox$3000 E, the corresponding error is 0.3 mrad with an uncertainty smaller than 0.1 mrad.

The final measured value for the WEP violation signal is 1.3(4.1) mrad, corresponding to a value of $0.9(2.9)\times 10^{-11}$ for the E$\rm \ddot o$tv$\rm \ddot o$s parameter, when $k_m^{\rm eff}$=$1.44(1)\times 10^9$ $\rm m^{-1}$, $\left(\frac{m_{\rm g}}{m_{\rm i}}\right)g$=9.793(1) $\rm{m/s^2}$, $T$=100 ms are substituted in Eq. (\ref{commonphase}).
For the evaluation of the systematic errors, many disturbances significant in usual gravity measurements are suppressed here by the common-mode measurement. We note that benefiting from the de-Broglie wave to enhance WEP test signal, a precision of 4.1 mrad for resolving the phase shift of the interferometer will lead to a test precision of $2.9\times 10^{-11}$ for our WEP test. By comparison, a precision of 4.1 mrad for resolving the phase shift corresponds only to a precision of 2.5 $\mu$Gal for usual gravity measurement with $T$=100 ms. On the other hand, phase shifts dependent on the initial velocity of the atom are usually avoided \cite{PhysRevA.93.023617,PhysRevLett.120.183604}, since the initial velocity is difficult to be measured or controlled with a high precision. However, WEP tests are null test experiments, and $(r_2-r_1)$ only needs to be measured with few significant figures. Thus, as the coefficient of $(r_2-r_1)$ in the test signal, $V_{\pi}$ only needs to be determined with a relative precision of about 10$\%$. For the corrections due to gravity gradient, FLS, etc., a precision of 1 mm/s for $V_{\pi}$ here is sufficient for our aimed test precision.

Our scheme is to incorporate the WEP test for atoms with respect to different HSs in AIs with HSs varied during interference. As showed in the typical MZI studied here, this incorporation yields a direct WEP test signal in single AI. With the idea of this incorporation, it is worthy to examine other types of AIs (for example, Ramsey interferometers \cite{PhysRev.78.695}, Ramsey-Bord\'{e} interferometers \cite{PhysRevA.30.1836} and AIs with multiple HSs involved) and the off-diagonal elements of $\widehat{M}_{\rm g} \widehat{M}_{\rm i}^{-1}$ as explored in Ref. \cite{Rosi_2017}. The WEP test signal is proportional to the de-Broglie wavelength in this situation, which means a possibility of WEP test using AIs insensitive to the absolute gravity, but sensitive to difference of the interested free fall accelerations \cite{PhysRevA.84.013620}. 
It is also expected that the WEP test using this scheme can be improved by a magnitude of two orders in a 10 m atomic fountain because of a matter-wave wave vector with a larger magnitude and a prolonged interrogation time \cite{PhysRevLett.111.083001,GRG10.1007}.
In conclusion, we demonstrate a de-Broglie wavelength enhanced WEP test for atoms in different HSs, and a violation of WEP is not observed at the level of $2.9\times10^{-11}$. This is the first WEP test using the de-Broglie wavelength, which effectively improves the test precision. This work provides an instructive idea for WEP tests with AIs.
 
The authors acknowledge the inspiring discussion with Shun Wang and Ze-Huang Lu. The authors also acknowledge anonymous referees for the insightful comments, which helped improve this work substantially.
The work was supported by the National Natural Science Foundation of China (Grants {No. 11727809}, {No. 11625417}, {No. 11904114} and {No. 11574099}).

\bibliographystyle{apsrev4-2}
\bibliography{Ref}

\end{document}